\documentclass[twocolumn,showpacs,amssymb,aps]{revtex4}

\usepackage{graphicx}
\usepackage{psfig}
\usepackage{dcolumn}
\usepackage{bm}

\def\lessim{\lower.5ex\hbox{$\; \buildrel < \over \sim \;$}}
\def\gtrsim{\lower.5ex\hbox{$\; \buildrel > \over \sim \;$}}
\begin{document} \hbadness=10000
\topmargin -0.8cm\oddsidemargin = -0.7cm\evensidemargin = -0.7cm
\preprint{}

\title{Centrality Dependence of Bulk Fireball Properties  at RHIC}
\author{Johann Rafelski}
\affiliation{Department of Physics, University of Arizona, Tucson, Arizona, 85721, USA}
\author{Jean Letessier}
\affiliation{Laboratoire de Physique Th\'eorique et Hautes Energies\\
Universit\'e Paris 7, 2 place Jussieu, F--75251 Cedex 05
}
%\\ \phantom{.}\\and}
\author{Giorgio Torrieri}%,
\affiliation{Department of Physics, McGill University, Montreal, QC H3A-2T8, Canada
}

\date{December 17, 2004}

\begin{abstract}
We explore the  centrality dependence  of  the properties of the dense hadronic
matter created in {$\sqrt{s_{NN}}=200$}\,GeV Au--Au collisions at RHIC.
Using the  statistical hadronization model,   we  fit   
particle yields known  for  11 centrality bins. 
We present the resulting model 
parameters, rapidity yields of physical quantities,
  and the physical properties of bulk matter at
 hadronization as function of centrality.
We discuss  the production of strangeness and entropy.
\end{abstract}

\maketitle
%%%%%%%%%%%%%%%%%%%
\section{Introduction}\label{Intro}
%%%%%%%%%%%%%%%%%%%%%%%%%%%%%%%%%%%%%%%%%%%%%%para 1
The  measurement of  hadron rapidity  yields 
at the Relativistic Heavy Ion 
Collider (RHIC) facilitates a study  of the physical 
properties of the hadronic fireball   at time 
of hadronization ({\it i.e.}, when these particles 
are produced). The objective of this work is to understand the impact parameter (reaction volume) 
dependence of the fireball bulk properties.
 We search for a  change of the reaction mechanism as
function of centrality: if
a new state of matter is formed in  central and semi-central 
 nuclear $AA$ collisions, but not in $pp$ reactions,
one would naively expect a visible change in some physical 
bulk properties for a sufficiently small number of reaction 
participants.

 We consider, at 
 the top RHIC energy $\sqrt{s_{NN}}=200$\,GeV Au--Au, 
the 11 centrality bins in which the 
$\pi^\pm, {\rm K}^\pm, p$ and $\bar p$ rapidity yields have been recently 
presented, see table I and table VIII in Ref. \cite{phenixyield}. 
These precise experimental  results involving a full range of centrality 
 motivate   this effort. 
We wish to establish, at the high level of precision now available
for the  RHIC $\sqrt{s_{\rm NN}}=200$ GeV run,
what a rapid change of the particle ratios such 
as $K^+/\pi^+$,  $K^-/\pi^-$  as function of centrality
 means both for the bulk physical  properties of the fireball,
and for the 
statistical hadronization model (SHM)
 parameters dependence on centrality.
 
These six particle rapidity yield results  
are complemented  with   STAR  results for the ratios 
${\rm K}^*(892)/{\rm K}^-$ \cite{haibin2200},
and  $\phi/{\rm K}^-$~\cite{phiyld}. 
Both ${\rm K}^*(892)/{\rm K}^-$ and  $\phi/{\rm K}^-$  do 
not show a  large  centrality dependence, but we make 
an effort to account for any dependence in our  analysis.

\begin{figure}[!bt]
%\hspace*{3cm}
%\vskip -0.5cm
%\hspace*{-.6cm}
\psfig{width=8.5cm,clip=,figure=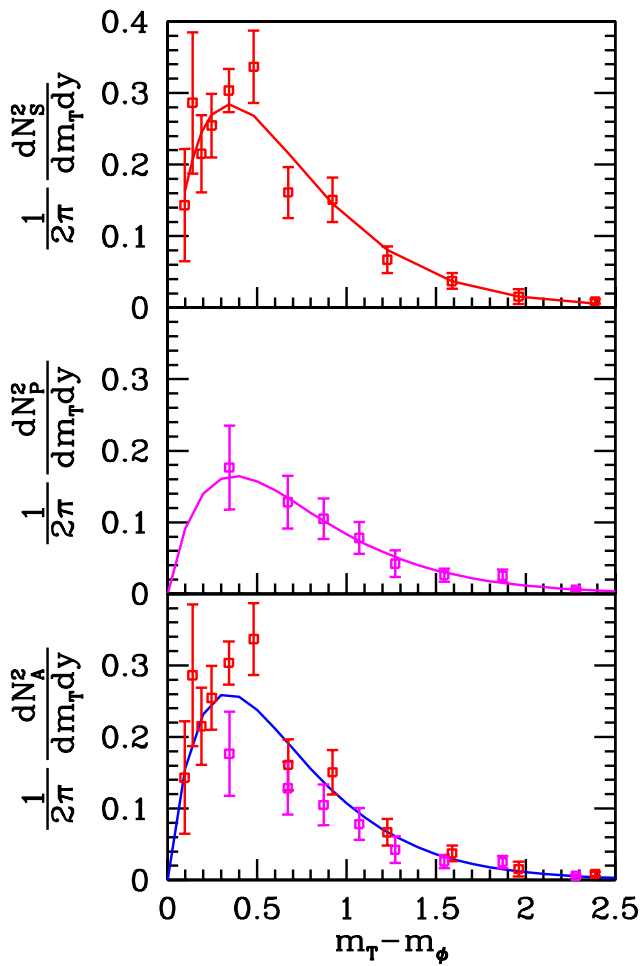}
%\vspace*{-0.6cm}
\caption{\label{phiSP}
(color online) Measured $\phi$-$m_\bot$ distributions
  $dN_i/dm_\bot dy$  for 10\% most 
central collisions at  $\sqrt{s_{NN}}=200$ GeV, lines are best fits.
From top to bottom: STAR (subscript $S$), PHENIX (subscript $P$)
and combined all data (subscript $A$). 
 }
%\vskip -0.3cm
\end{figure}

We considered  the difference between STAR~\cite{phiyld}, and    
PHENIX (submitted for publication, \cite{Adler:2004hv}) $\phi$-results. We illustrate the situation in Fig.\,\ref{phiSP}~\cite{phidata}. The lines  
 show our best fit results to STAR (top panel), PHENIX (middle panel) and
combined data set (bottom panel). The integrated yields agree  
 for the top two panels with those reported by the experimental
collaborations. We note  that the integrated yield derived from the 
combined data fit   (bottom panel of Fig.\,\ref{phiSP}), 
to all available 10\% centrality $\phi$-yields, is not compatible  
 with the PHENIX yield. This is so, since the evaluation of the integrated 
PHENIX   $\phi$-yield depends on   the lowest  $m_\bot$
measured  yield. This data point  appears to be a 1.5 s.d. low anomaly  
compared to the many  STAR $\phi$-results available at low~$m_\bot$.
This possibly statistical fluctuation materially influences the
total integrated PHENIX $\phi$-yield.

%%%%%%%%%%%%%%%%%%%%%%%%%%
%%%%%%%%%%%%%%%%%%%%%%%%%%

 The ratio ${\rm K}^*(892)/{\rm K}^-$ 
anchors and confirms the chemical freeze-out 
temperature $T$, which is the only parameter on which this ratio 
depends. The ratio $\phi/{\rm K}^-$ 
 comprises a multi-strange particle and
anchors and confirms the chemical conditions at freeze-out.
For this reason use of these  yield results is of essence to obtain 
the precision results we present here. However,  
in principle the resonance yields maybe  significantly altered  
by post-hadronization  processes~\cite{Bleicher,Fachini,Zhangbu}, 
or their  observable yield could be 
 impacted by decay product rescattering~\cite{TorrRes}. 

The  chemical
non-equilibrium hadronization  model   describes the experimental
data analyzed very well. Moreover, in this model, one 
expects   near coincidence
of the thermal and chemical freeze-out. In this limit,
there is no post-hadronization resonance yield evolution, or
significant decay product rescattering. We have never
come across the   need for SHM adjustments of yields of the 
K$^*$-resonance or $\phi$. Therefore, we do not pursue the 
development of     kinetic   yield evolution  models for these 
particles.
The interested reader can 
follow up these developments in Refs.\,\cite{Bleicher,Fachini,Zhangbu}.

%%%%%%%%%%%%%%%%%%%
\section{Statistical Hadronization Models and Statistical Parameters}
\label{SHMsec}
%%%%%%%%%%%%%%%%%%%%%%%%%%%%%%%%%%%%%%%%%%%%%%para 1
The statistical hadronization model  is, by definition, a model  of 
particle production in which the birth process of each particle 
fully saturates (maximizes) the quantum mechanical probability amplitude, 
and thus, the yields are determined by the appropriate integrals of the
accessible phase space \cite{JJBook}.
 For a system subject to global dynamical 
evolution, such as collective flow,  this is understood to apply 
within each local  co-moving frame element. The results presented
here were obtained using the numerical package SHARE 
(Statistical Hadronization with REsonances) \cite{share}. 

The question, if SHM  
is indeed consistent with the wealth of RHIC data available today,
comes to mind. Our comprehensive study of central reactions 
at central rapidity, for both  $\sqrt{s_{\rm NN}}=130$ and 200~GeV,  
suggests so strongly \cite{Letessier:2005qe}.
Systematic study of particle production for 
a wide reaction energy range confirms applicability of 
the SHM, for review see  \cite{BDMRHIC}.  
At RHIC there are   two 2 s.d. (standard deviation) 
  exceptions among the model agreement with  
hadron  particle yields at RHIC-200:\\
1) the  (preliminary) $\overline{\Omega}/\Omega=1.01\pm0.08$ 
yield ratio \cite{barannikova}, with the central value  
 greater than unity 
while, on general grounds, at finite baryon density 
this ratio should be smaller than unity;\\
 2)  the  (preliminary) $\Delta^{++}/p=0.24\pm 0.06$ 
 \cite{Markert,barannikova,haibin2200}, which 
in statistical hadronization
models  is half as large. We note that 
 the central value of this   result means  that,  after removal
 of  descendants from weak decays,  nearly
all protons observed  should have been a primary  $\Delta$.

In addition to the 8 particle (relative) yields considered, we also enforce 
three  supplemental constraints:\\
 1) strangeness conservation, {\it i.e.}, the (grand canonical)
 count of $s$ quarks in all hadrons equals such  $\bar s$ count for 
each rapidity unit;\\
 2) the electrical charge to net baryon ratio 
in the final state is the same as in the initial state to within 2\%;\\
 3) the ratio   $\pi^+/\pi^-=1.\pm0.02$, which helps constrain the 
isospin asymmetry.\\
This last ratio appears redundant, as we already independently  use
the yields of $\pi^+$ and $\pi^-$. These yields have a large systematic
error and do not constrain their ratio well,
 and thus the supplemental constraint
is introduced, since SHARE allows for the isospin asymmetry effect.

The successful description of  particle yields 
within the SHM obtained for a single chemical
freeze-out condition produces, as a first result,  the model 
parameters in the process of $\chi^2$ minimization:  the  (chemical)
freeze-out temperature $T$, the baryon   $ \mu_B$ and hyperon $\mu_S$ 
chemical potentials. We obtain and present  results  at three chemical condition
alternatives, the chemical equilibrium (dashed lines, red online), 
strange quark non-equilibrium with 
phase space occupancy $\gamma_s\ne 1$ (dotted lines, violet online), 
and  the light quark  
flavor yield  (full) non-equilibrium model including  $\gamma_q\ne 1$
(solid lines, blue online).
In our approach,  4 to 6 parameters confront, in a systematic fashion, 
11 yields and/or ratios and/or constraints containing one redundancy. 

\begin{figure}[!bt]
%\hspace*{3cm}
%\vskip -0.5cm
\hspace*{-.6cm}
\psfig{width=8.5cm,figure=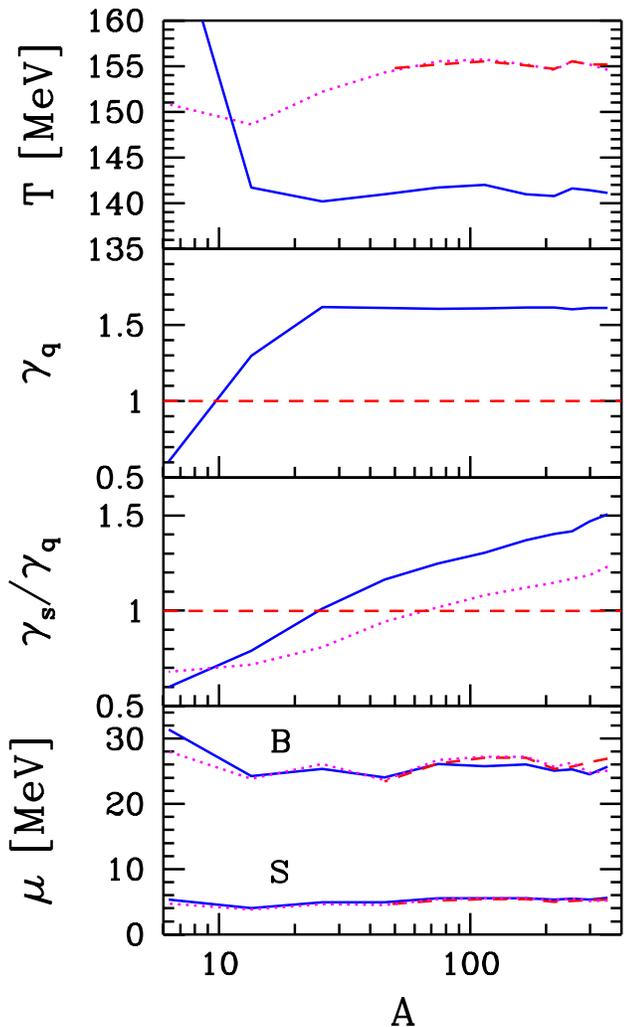}
%\vspace*{-0.6cm}
\caption{\label{TGMU}
(color online) From top to bottom: temperature $T$, light quark phase 
space occupancy $\gamma_q$,
the ratio  of strange to light quark phase 
space occupancies $\gamma_s/\gamma_q$  and the chemical potentials
($B$ for baryochemical $\mu_B$ and $S$ for strangeness  $\mu_S$) 
as function of centrality (average participant number  $A$). 
The lines connect the results obtained at each bin center $A$, 
and have not been smoothed in order to show result fluctuations:
1)~Full chemical non-equilibrium  model ---  
(blue online)  solid lines; 
2)~Strangeness chemical non-equilibrium  model ---  
(violet online) dotted lines; 
3)~Chemical equilibrium model --- 
(red online)   dashed lines.
  }
%\vskip -0.3cm
\end{figure}

The results we present for the model parameters,  in Fig.\,\ref{TGMU},  have all 
  above 85\% confidence level (we do not present the low
centrality chemical equilibrium results as these do not satisfy this 
criterion). In table \ref{ResTable}, we present all these results 
with precision which should help reproduce particle yields when required.
In general, the reader should not expect to reproduce the last fourth digit shown.
We also present, in table \ref{ResTable}, along with $A$, and in same precision, 
the volume normalization  factor $dV/dy$, which is required to obtain 
the particle yields. We further note that only one of the three models 
presented is applicable, and thus, the variation of parameters between these 
models must be seen as the sensitivity  of the data  to 
 their physical relevance. 

%%%%%%%%%%%%%%%%%%%%%%%%%%
%%%%%%%%%%%%%%%%%%%%%%%%%%

%%%%%%%%%%%%%%%%%%%%%%%%%%%%%%%%%%
\begin{table}[bt]
\caption{
\label{ResTable}
Fitted statistical parameters, for each central value of centrality expressed in 
terms of participant number $A$, as defined by PHENIX; for the three models in sequence: 
chemical non-equilibrium model, chemical semi-equilibrium, and chemical
equilibrium. For $dV/dy$ the unit is fm$^3$, for $T,\mu_{\rm B}, \mu_{\rm S}$ the unit is  MeV, 
all other quantities are dimensionless.
}\vspace*{0.2cm}%\large%\small
\begin{tabular}{|c c|| c| c | c| c|| c| c|}
\hline
$A$   &$dV/dy$& $T$   & $\mu_{\rm B}$  &  $\mu_{\rm S}$  &\ $\lambda_{I3}$\   &\   $\gamma_s$\   &  \  $\gamma_q$\  \\
\hline
351.4 &  969  & 141.1 &  25.67 &  5.592 & 0.9967  &  2.430 & 1.613\\
299.0 &  821  & 141.4 &  24.52 &  5.34\ & 0.9969  &  2.367 & 1.612\\
253.9 &  706  & 141.6 &  25.27 &  5.463 & 0.9968  &  2.270 & 1.603\\
215.3 &  611  & 140.8 &  25.05 &  5.325 & 0.9969  &  2.266 & 1.615\\
166.6 &  462  & 141.0 &  26.01 &  5.523 & 0.9968  &  2.212 & 1.614\\
114.2 &  298  & 142.0 &  25.75 &  5.528 & 0.9968  &  2.096 & 1.608\\
 74.4 &  192  & 141.7 &  26.14 &  5.518 & 0.9968  &  2.003 & 1.605\\
 45.5 &  119  & 141.0 &  24.05 &  4.929 & 0.9972  &  1.876 & 1.613\\
 25.7 &  68.1 & 140.2 &  25.32 &  4.953 & 0.9972  &  1.636 & 1.618\\ 
 13.4 &  55.1 & 141.7 &  24.24 &  4.045 & 0.9970  &  1.026 & 1.299\\
  6.3 &  42.6 & 172.7 &  31.39 &  5.356 & 0.9954  &  0.363 & 0.606\\
\hline
\hline
351.4 & 1735 & 154.6  &25.04  & 5.161   & 0.9958  &  1.231  & 1\\
299.0 & 1458 & 155.2  &24.73 &  5.110   & 0.9958  &  1.186  & 1\\
253.9 & 1215 & 155.5  &26.29 &  5.441   &0.9956   &  1.169  & 1\\
215.3 & 1072 & 154.6  &25.68 &  5.206  & 0.9957   &  1.147  & 1\\
166.6 &  795 & 155.2 & 27.18 &  5.540  & 0.9955   &  1.121  & 1\\
114.2 &  521 & 155.7 & 27.21 &  5.555   & 0.9955  &  1.080  & 1\\
 74.4 &  334 & 155.5 & 26.74 &  5.367   & 0.9956  &  1.018  & 1\\
 45.5 &  241 & 152.6 & 21.62 &  3.972   & 0.9967  &  0.8906  & 1\\
 25.7 &  131 & 152.2 & 26.12  & 4.661   & 0.9962  &  0.8076  & 1\\
 13.4 &  74.9 & 148.6 & 23.82  & 3.821   & 0.9969  &  0.7163  & 1\\
  6.3 &  34.7 & 150.8  &28.00  & 4.681   & 0.9961  &  0.6788  & 1\\
\hline
\hline
351.4 &  1920 & 155.2  & 26.93 &  5.349  & 0.9956  &  1  &  1 \\
299.0 &  1609 & 155.2 &  26.41 &  5.249  & 0.9957  &  1  &  1 \\
253.9  & 1328  &155.5 &  25.70 &  5.137  & 0.9958  &  1   & 1 \\
215.3  & 1157 & 154.7 &  25.32  & 4.982  & 0.9959  &  1   & 1 \\ 
166.6 &   855 & 155.1 &  27.06 &  5.369  & 0.9956  &  1   & 1 \\ 
114.2 &   550 & 155.5  & 27.06 &  5.414  & 0.9956  &  1   & 1 \\ 
 74.4 &   342 & 155.2  & 26.15 &  5.200  & 0.9957  &  1   &  1 \\
% 45.5  &  213 & 153.6 & 22.20 &  4.266   & 0.9965 &  1  &  1 \\
% 25.7 &   105 & 153.0  & 34.84  & 6.628   & 0.9946 &  1  &  1 \\ 
% 13.4 &   57.9 & 149.2 &  20.85  & 3.27   & 0.9977  &  1  &  1 \\ 
%  6.3 &   25.1 & 151.9 &  24.84  & 4.60   & 0.9962  &  1  &  1 \\ 
\hline
\hline
\end{tabular}\vspace*{0.1cm}
 \end{table}
%%%%%%%%%%%%%%%%%%%%%%%%%%%%%%%%%%%%%%%%%%%%%%%%%%%%%%%%%%%%%%%%%%%%%%%%%%%%%%%%%%%%%

A notable feature, in  Fig.\,\ref{TGMU}, is  absence of 
centrality features in temperature $T$ and chemical potentials $\mu_{B,S}$,
except for the most peripheral centrality,  and up to the variation 
which can be associated with fluctuation
in the data sample and/or determination of the minimum of $\chi^2$. 
The deviation,  at  the most peripheral centrality bin
from trends set by other results, could be an indication of 
the change in the reaction mechanism for which we are looking.

Independent of the chemical (non-)equilibrium
assumption, the baryochemical potential   $\mu_B=25\pm1$ MeV across the
10 centrality bins. Similarly,  we find strangeness  chemical potential
 $\mu_S=5.5\pm0.5$ MeV (related to   strange quark chemical potential
$\mu_s=\mu_B/3-\mu_S$). The freeze-out temperature is for the 
semi-equilibrium and equilibrium model about  10\% greater than 
the full chemical non-equilibrium freeze-out. 
The most notable  variation,  in  Fig.\,\ref{TGMU}, is 
the gradual increase in strangeness phase space 
occupancy $\gamma_s/\gamma_q$  and thus strangeness yield 
with collision centrality.   This effect was predicted and  
originates in an increasing  lifespan of the fireball \cite{impact}.
The over-saturation of the phase space has been also
 expected  due to both, the dynamics of expansion \cite{RHICPred},  
and/or reduction in phase space size as a parton based matter turns
into HG \cite{JRBielefeld}. This latter
effect is also held responsible for the saturation
of light quark phase space $\gamma_q\to e^{m_\pi/2T}$. 
A systematic increase of $\gamma_s$ with collision centrality 
has been reported for several reaction energies  \cite{Kampfer:2003pf}.

Using SHARE, we find  $T=155\pm8$ MeV for the 
 chemical equilibrium and strangeness non-equilibrium
freeze-out. The error is our estimate of the propagation 
of the systematic data error, combined with the fit uncertainty;
the reader should note that the   error comparing centrality
to centrality is negligible. This result for $T$  is in mild
disagreement (1.5 s.d.) with earlier chemical  equilibrium 
fits  \cite{BDMRHIC,Broniowski:2003ax}. 
This, we believe, is due  to: 
1) differences in data sample used, specifically,
the  hadron resonance production results used
provide  a very strong constraint for the fitted 
temperature, and 2) the more  complete 
treatment by  SHARE of  hadron mass spectrum and of heavy
resonances multi-particle decays.

We find two regimes of hadronization temperature:\\
a) for the chemical equilibrium case, and for the
 chemical semi-equilibrium case
the hadronization temperature $T=155\pm8$ MeV   is right at the 
value of the cross-over temperature for   
 lattice QCD with 2+1 dynamical flavors results for 
$\mu_{\rm B}\to 0$ in Refs.\,\cite{lattice2} and \cite{Fodor:2004nz}.
These two groups  apply different methods and approximations, their results
imply that at RHIC conditions the temperature of the cross-over from
QGP to HG in chemical equilibrium is near to $T=160\pm5$ MeV.\\ 
b) The chemical non-equilibrium model requires for internal consistency
a fast hadronization process, for which the sole known mechanism 
is  super cooling of the fireball due to  rapid transverse 
expansion~\cite{suddenPRL}. 
Supercooling implies that the  hadronization temperature is below
phase cross-over boundary, and the value  $T=141\pm7$ MeV, we have found 
in the full chemical non-equilibrium 
fit, is a result internally consistent with this reaction mechanism.
These issues are   discussed further in the context of the search of 
phase threshold as function of reaction energy, see Ref.\,\cite{Letessier:2005qe}.

For the chemical non-equilibrium sudden hadronization reaction 
picture, we further expect that the hadronization geometry is
highly non-homogeneous, corresponding to, {\it e.g.}, fingering  breakup of the fireball.
This makes it possible that the thermal and chemical freeze-out condition 
coincide~\cite{Torrieri:2000xi,Broniowski:2001we}, as the surface
to volume area is large, emitted particles having a small probability to 
rescatter. For this reason, as noted before,   there
is no   need to discuss changes in observable 
hadron resonance populations 
which may arise in post-hadronization scattering   processes.

%%%%%%%%%%%%%%%%%%%
\section{Physical Properties of the Fireball at Hadronization}
\label{PhysProp}
%%%%%%%%%%%%%%%%%%%%%%%%%%%%%%%%%%%%%%%%%%%%% 

%%%%%%%%%%%%%%%%%%%%%%%%%%
\begin{figure}[!bt]
%\hspace*{3cm}
%\vskip -0.5cm
\hspace*{-.6cm}\psfig{width=8.5cm,figure=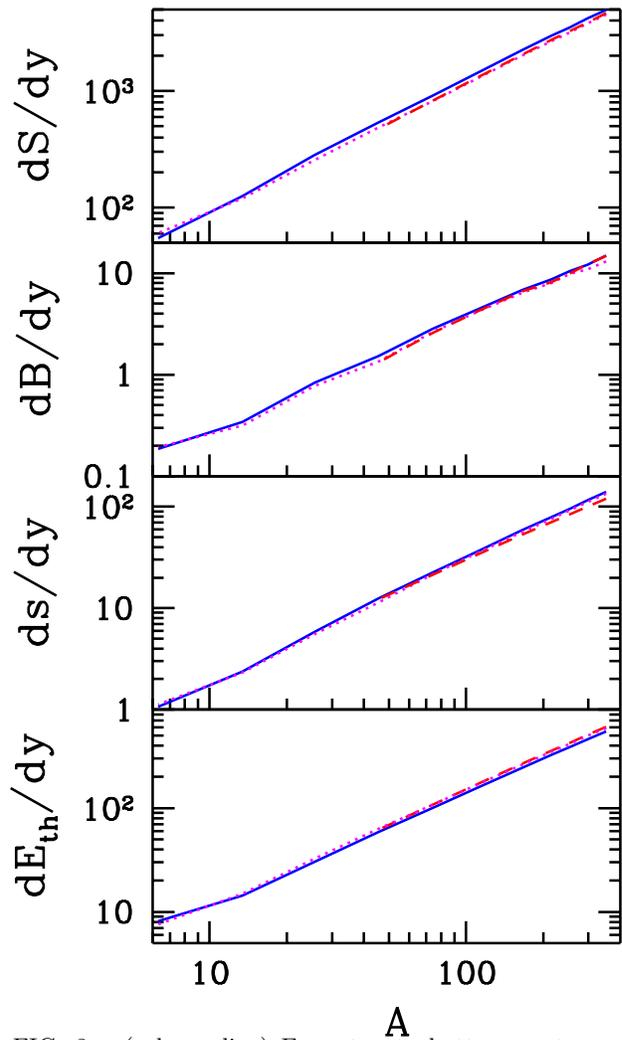}
\vspace*{-0.6cm}
\caption{\label{yyields}
(color online)  From top to bottom: entropy, net baryon, 
strangeness and thermal energy 
yield  per unit of rapidity, as a function of centrality. 
 Lines (nearly overlapping) are coded as in   figure \ref{TGMU} for
the three chemical models.
 }
%\vskip -0.3cm
\end{figure}
%%%%%%%%%%%%%%%%%%%%%%%%%%
 
Given the statistical parameters, we can evaluate the 
yields of particles not yet measured and obtain
the rapidity yields of  entropy, net baryon number, net strangeness,
and thermal energy as function of  centrality,
shown  in Fig.\,\ref{yyields}.  We find, in the most central reaction bin,
  $14.9\pm 1.5$ baryons per unit rapidity interval, a rather 
large baryon stopping in the central rapidity domain. 

The  entropy yield is reaching $dS/dy= 5000\pm 500$. This compares to
 the estimate made recently by Pal and Pratt 
\cite{Pal:2003rz} who find $dS/dy=4451\pm 445$ for the most central 
130 GeV  reactions. One should note that the smoothness of the results,
presented in Fig.\,\ref{yyields} as function of $A$, is directly related 
to the smoothness of experimental data, which determines in some cases 
nearly directly  these observables. For example,
 the hadron multiplicity per rapidity unit 
is directly related to the  entropy yield. Thus, the experimental yield error
of 10\% is directly the error of entropy $dS/dy$. If instead, we constructed this
error from partial errors in the statistical parameters,  the implicit
cancellations would be hard to realize. For this reason, we do not 
state these errors in Fig.\,\ref{yyields}, all these micro canonical 
quantities are directly relate to particle yields and their
error is at 10\% level.

%%%%%%%%%%%%%%%%%%%%%%%%%%
\begin{figure}[!bt]
%\hspace*{3cm}
%\vskip -0.5cm
\hspace*{-.6cm}\psfig{width=8.5cm,figure=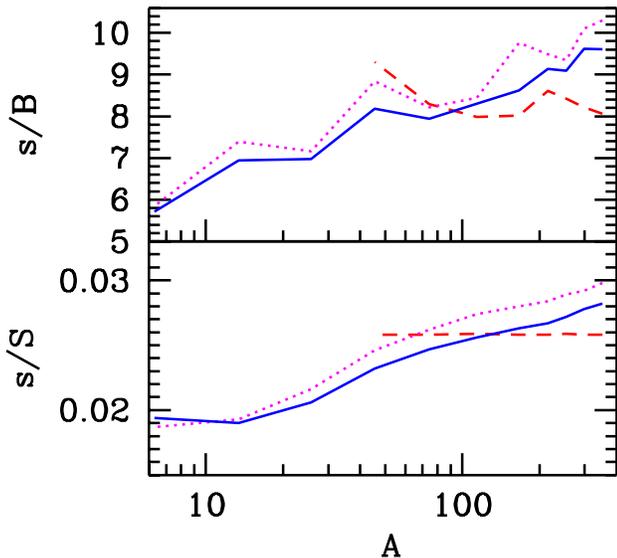}
\vspace*{-0.6cm}
\caption{\label{SBS}
(color online)  From top to bottom : strangeness per net baryon $s/B$ and 
strangeness per entropy $s/S$, as a function of centrality.
 Lines  are coded as in   figure \ref{TGMU} for
the three chemical models.
  }
%\vskip -0.3cm
\end{figure}
%%%%%%%%%%%%%%%%%%%%%%%%%%
 
Except for the chemical  equilibrium  model (which has a restricted 
centrality range of validity, for it fails to fit the peripheral collisions), 
the rise of strangeness yield with 
centrality is faster than the rise of baryon number yield:
 $(ds/dy)/ (dB/dy)\equiv s/B$ is seen in  the top panel of Fig.\,\ref{SBS}.
For the  most central reactions, we  reach  $ s/B =9.6\pm 1$. We also note
 fluctuations in the trend of the results of 
 the magnitude expected from the experimental data error. 

The  increase, with $A$, of per baryon specific strangeness yield 
indicates presence of a  production  mechanism acting beyond the first collision
dynamics. This new mechanism must  benefit  from the increased 
size, or more appropriately, increased life span of the larger reaction system.
We have shown earlier that the thermal gluon fusion to strangeness will 
have just this behavior~\cite{impact}.

While  entropy production occurs 
predominantly   during  the initial parton thermalization phase,  
thermal strangeness production requires presence of thermal, mobile
gluons, being   driven by  thermal  gluon fusion~\cite{RM82}. Thus,
strangeness production follows in time the entropy production, and is 
strongest at the highest available temperature considering  
the strangeness mass threshold. 

To understand this better, we show, in the   bottom panel  of Fig.\,\ref{SBS},
strangeness per  entropy  $s/S$, as function of centrality. We find a 
smooth transition from a flat peripheral  behavior where $s/S\lessim 0.02$, 
to a smoothly increasing $s/S$  reaching $s/S\simeq 0.03$ for most  central 
reactions. This rise, occurring  for  $A>20$, indicates the onset of an 
additional strangeness production mechanism. 

%\floatfix

  As the system size increases, the 
time span, during  which thermal strangeness production is effective, is 
increasing allowing  strangeness to   approach  chemical equilibration 
in the parton phase.  A back of envelope estimate of the ratio of $s/S$, 
in a chemically equilibrated parton plasma, yields the value seen in 
Fig.\,\ref{SBS}  for large $A$~\cite{Letessier:2005qe}. 
We thus understand the rise and magnitude 
of  $s/S$  ratio in terms of the expected reaction 
mechanisms in the deconfined 
phase, and the chemical saturation of this ratio. Chemical 
saturation in QGP, {\it i.e.\/}, $\gamma_s^{\rm QGP}\to 1$, 
implies, due to higher strangeness phase space content in QGP than 
in hadron matter, that $\gamma_s>1$.

%%%%%%%%%%%%%%%%%%%%%%%%%%
\begin{figure}[!bt]
%\hspace*{3cm}
%\vskip -0.5cm
\hspace*{-.6cm}\psfig{width=8.5cm,figure=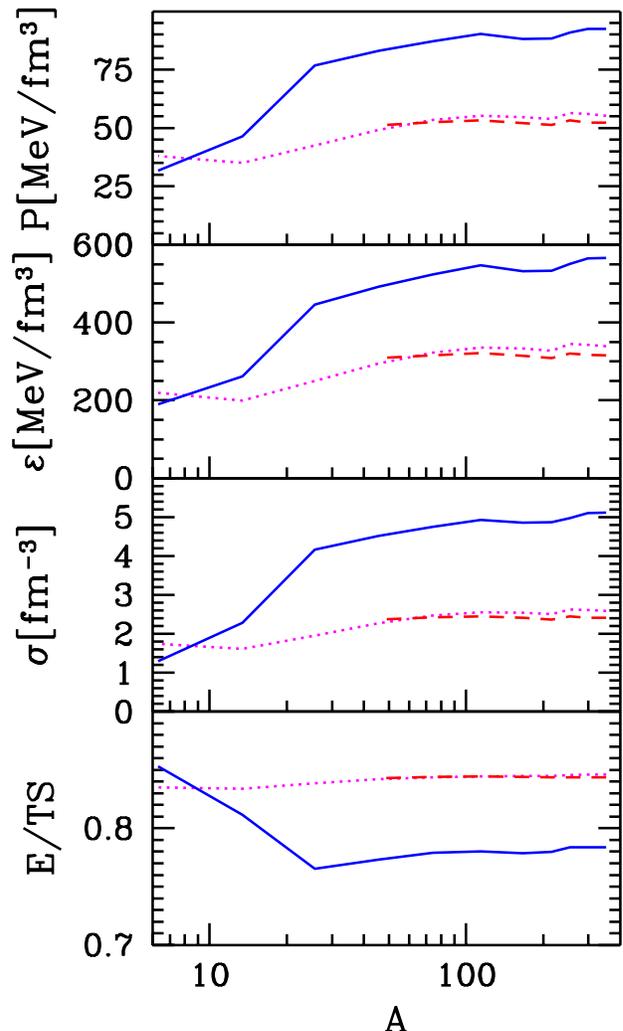}
\vspace*{-0.6cm}
\caption{\label{PEST}
(color online) From top to bottom:  pressure $P$, energy density $\epsilon=E/V$, 
entropy density $S/V$ and   $E/TS$,  
as a function of centrality.  Lines  are coded as in   figure \ref{TGMU} for
the three chemical models.
}%\vskip -0.3cm
\end{figure}
%%%%%%%%%%%%%%%%%%%%%%%%%%

 We evaluate now the bulk  properties of the fireball  
shown  in Fig.\,\ref{PEST}.
To obtain the pressure $P$, the 
energy density $\varepsilon\equiv (dE/dy)\,/(dV/dy)$
and the entropy density $\sigma\equiv (dS/dy)\,/(dV/dy)$,
 we sum, using the SHARE data base for hadron 
resonances,  all partial particle  
contributions using relativistic expression, see \cite{JJBook}.
 Since we fitted six  particle rapidity  yields,  the global 
fitted yield normalization factor $dV(A)/dy $  
 is reliable, up to the systematic
error of about 10\% inherent in the PHENIX rapidity 
yield data. 

When we consider ratios of two 
bulk properties, {\it e.g.},  $E/TS$, the  data and fit fluctuations cancel out, 
and  the results are in general much smoother. Moreover, 
such ratios do not depend on the identification in  the  normalization 
factor $dV/dy$ with the statistical 
model volume. For example, when one considers hadrons of finite proper
 volume \cite{HR}, there is a  correction factor.
Hence, the smooth and precise values for  $E/TS$ which are not expected
to change in their interpretation offer a  challenge for 
structure model of hadronization, and  will need to be
addressed quantitatively. We note, as an example, that  a quark matter
system consisting of thermal mass quarks with $m=aT$, $a\simeq 2$ for
non-equilibrium and $a=4$ for equilibrium,  will
yield just the result seen in the bottom panel of Fig.\,\ref{PEST}. 

We note, in Fig.\,\ref{PEST}, that the chemical non-equilibrium system  
is much denser at hadronization \cite{Bari}. 
The entropy density, assuming chemical 
non-equilibrium, is   by more than a factor 2 larger compared 
to the chemical equilibrium model. It is
for this reason that the ratio $E/TS$, in Fig.\,\ref{PEST}, is
for the non-equilibrium case significantly smaller than it is when assuming
equilibrium. 

%%%%%%%%%%%%%%%%%%%%%%%%%%%%%%%%%%
\section{Final Remarks}
%%%%%%%%%%%%%%%%%%%%%%%%
We have presented a comprehensive analysis of soft hadron  
yields   at  $\sqrt{s_{\rm NN}}=200$ GeV as function of centrality. We have 
obtained the statistical hadronization 
parameters that describe the data, and evaluated
the physical properties of the hot fireball at hadronization. Our analysis
included, aside of `stable' hadron PHENIX data ($\pi^\pm,\,{\rm K}^\pm, p,\bar p$),
also the STAR $K^*$ and $\phi$ yields. For the latter, we also presented
a comparison analysis in order to resolve a discrepancy between STAR and PHENIX
$\phi$-yield result. We found that, for $A\gtrsim 20$, the  statistical parameters
do not vary with centrality, with the exception of strangeness 
quark occupancy, $\gamma_s$. Already for $A\gtrsim 20$, the system properties
are approaching those seen for the greatest available $A\simeq 350$. 
However, for $A<20$ there is a significant change in the physical 
and statistical properties of the fireball. The chemical equilibrium
description of the experimental  results here considered 
is not possible for these small fireballs. 

We are aware of  two prior efforts to explore  statistical parameters,
 but not bulk properties, as function of collision centrality. 
At  $\sqrt{s_{\rm NN}}=130$ GeV \cite{Cleymans:2004pp}, 
the analysis remains inconclusive in view of  the limitations of 
  the experimental data. At  $\sqrt{s_{\rm NN}}=200$ GeV,  an analysis
  assuming chemical semi-equilibrium~\cite{Kaneta:2004zr} shows
 trends  comparable with those we found.   However, our hadronization
temperature $T$ and phase space occupancy $\gamma_s$ 
are  anchored by  the ${\rm K}^*(892)$ and $\phi$ 
yields, and hence, we obtain for the chemical semi-equilibrium a  smaller  
 value of $T$, and therefore a greater $\gamma_s$ which
for most central reactions clearly exceeds unity. 

In this work, we have obtained the centrality dependence
of the  hadronization   pressure, entropy density  and thermal
 energy density, which are  
  of the magnitude expected:  for the chemical non-equilibrium model  
the pressure saturates at $P=92$ MeV/fm$^3$, a typical 
value we are used to from the bag model of hadrons to be the vacuum 
pressure.  The energy density of 0.5 GeV/fm$^3$ is in accord with 
lattice results    for the energy density of matter 
subject to the phase transformation to/from deconfinement.

We have shown that the fireball strangeness content is increasing fastest 
with increasing  centrality, beating out in the competition both, 
the stopped net baryon number, and the produced entropy. 
We have shown that the ratio $s/S$ saturates near the 
value of chemically equilibrated QGP phase.

In our opinion, the most remarkable finding of this study is the recognition
that the statistical parameters, and thus also the
 bulk properties of dense matter fireball created 
at RHIC, do not depend on the size of the system 
for $A\gtrsim 20 $, where $A$ is the number of reaction participants.
There is a rapid adjustment in the value of statistical parameters fitted:
  temperature is dropping and $\gamma_q$ is increasing, which changes combine
to yield a doubling in the density of the hadronizing fireball for $A\gtrsim 20 $.
Interpreting the high density phase at hadronization as the deconfined 
state, our results can be interpreted to indicate that  at
RHIC energy scale the quark liquid phase is formed for  $A\gtrsim 20 $.

%\vspace*{.5cm}
\acknowledgments
We thank Jingguo Ma   and   Zhangbu Xu for valuable comments, discussion,
and the suggestion that the PHENIX $\phi$-yield results are not inconsistent 
with STAR results when included in a common data analysis. 

Work supported by a grant from the U.S. Department of
Energy  DE-FG02-04ER41318, 
the Natural Sciences and Engineering research
council of Canada, the Fonds Nature et Technologies of Quebec. 
LPTHE, Univ.\,Paris 6 et 7 is: Unit\'e mixte de Recherche du CNRS, UMR7589.
G. T. thanks the Tomlinson foundation for support given.

%%%%%%%%%%%%%%%%%%%%%%%%%%%%%%%%%%%%%%%%%

%%%%%%%%%%%%%%%%%%%%%%%%%%%%%%%%%%%%%%%%%
\vskip 0.3cm
%%%%%%%%%%%%%%%%%%%%%%%%%%%%%%%%%%%%%%%%%
%\begin{references}

\end{document}